\documentclass{article}

\usepackage{arxiv}

\usepackage[utf8]{inputenc} 
\usepackage[T1]{fontenc}    
\usepackage{hyperref}       
\usepackage{url}            
\usepackage{booktabs}       
\usepackage{amsfonts}       
\usepackage{nicefrac}       
\usepackage{microtype}      
\usepackage{lipsum}		
\usepackage{graphicx}
\usepackage[numbers]{natbib}

\usepackage{doi}
\newtheorem{lemma}{Example}

\title{Methods for Eliciting Informative Prior Distributions: A Critical Review}

\author{ \href{https://orcid.org/
0000-0002-7970-6342}{Julia R. Falconer}\thanks{Julia Falconer's research is funded through the University of Waikato Doctoral Scholarship} \\
	Department of Mathematics,\\
	University of Waikato, \\
	Hamilton, New Zealand \\
	\texttt{jrg22@students.waikato.ac.nz} \\
	\And
	{Eibe Frank} \\
	Department of Computer Science,\\
	University of Waikato, \\
	Hamilton,New Zealand\\
	 \\
	\And
	{Devon L. L. Polaschek}\\
	School of Psychology,\\
	University of Waikato, \\
	Hamilton, New Zealand
	\And
	{Chaitanya Joshi}\\
	Department of Mathematics,\\
	University of Waikato, \\
	Hamilton, New Zealand \\
}

\renewcommand{\shorttitle}{\textit{Methods for Eliciting Informative Prior Distributions: A Critical Review}}

\hypersetup{
	pdftitle={Methods for Eliciting Informative Prior Distributions},
	pdfsubject={q-bio.NC, q-bio.QM},
	pdfauthor={Julia R. Falconer, Eibe Frank, Devon L. L. Polascheck, Chaitanya Joshi},
	pdfkeywords={Bayesian inference, prior elicitation, informative priors,  uncertainty},
}

\begin{document}
	\maketitle
	
	\begin{abstract}
		Eliciting informative prior distributions for Bayesian inference can often be complex and challenging. While popular methods rely on asking experts probability based questions to quantify uncertainty, these methods are not without their drawbacks and many alternative elicitation methods exist. This paper explores methods for eliciting informative priors categorized by type and briefly discusses their strengths and limitations. Most of the review literature in this field focuses on a particular type of elicitation approach. The primary aim of this work, however, is to provide a more complete yet macro view of the state of the art by highlighting new (and old) approaches in one clear, easy to read article. Two representative applications are used throughout to explore the suitability, or lack thereof, of the existing methods; one of which, highlights a challenge that has not been addressed in the literature yet. We identify some of the gaps in the present work and discuss directions for future research.
	\end{abstract}

	\keywords{Prior Elicitation \and Bayesian Inference \and Informative Priors \and Uncertainty}
	
	\section{Introduction}
	Bayesian inference can be thought of as the process of updating prior knowledge once the data, $y$, has been observed. Bayes' rule gives
	\begin{equation}
	p(\theta | y) \propto p(\theta)p(y|\theta),
	\end{equation}
	where $p(\theta)$ is the prior distribution on parameters of interest, $\theta$, and $p(y |  \theta)$ is the likelihood function for data, $y$. The prior distribution could be considered to be \textit{informative}, that is, elicited based on available information and beliefs, or \textit{non-informative}, where no such prior information or beliefs may be available.
	If there is a large amount of data then the inference will be influenced more by the likelihood function. The opposite of this is also true; if there is a limited amount of data, information from the likelihood will be weak and the inference will be influenced more by the prior distribution. In this case, the choice of the prior, $p(\theta)$ is of greater importance. Ideally, an informative prior would be used, because the decision making and inference for this problem will primarily rely on the prior.\\

	A simple example  of when an informative prior may be required is as follows,
	\begin{lemma}\label{example1}
		A pharmaceutical company wishes to predict whether or not patients will develop blood clots after taking a new drug they are developing. Let $Y \in \{1,0\}$, therefore $Y \sim Bernoulli(\theta)$, where $\theta$ is the probability of a patient developing blood clots after taking the new drug. Data on existing patients is limited due to the new formula of the molecule. The goal is to obtain a prior distribution on $\theta$ which can be used for further research.
	\end{lemma}

	For Example \ref{example1}, a common approach for obtaining an informative prior distribution would be to elicit information from an expert, i.e., to perform expert \textit{prior elicitation}. The process may require multiple individuals with different expertise and roles. A set of standard definitions of these, as used throughout this paper, can be found in Table \ref{Tabledef}.
	Although obtaining information from experts may seem straightforward at first glance, it often is anything but.  Several challenges may arise, complicating the process. We discuss the key challenges below.\\
	
	\begin{table}[h]
		\caption{Definitions}
		\centering
		\begin{tabular}{l  p{.5\textwidth}}
			\hline
			Name     & Description\\
			\hline
			\emph{Prior Elicitation} & This paper refers to prior elicitation as the process of obtaining knowledge from a source to form a prior distribution which can be used for further Bayesian analysis. Also referred to in texts as "Probability Encoding".     \\
			\hline
			\emph{Expert} & An expert is an individual who has extensive knowledge on a certain subject matter. Also referred to some in  texts as a "judge".\\ \hline
			\emph{Analyst} & An analyst is an individual who performs the task of forming a prior distribution using prior elicitation techniques. \\
			\hline
			\emph{Facilitator} & A facilitator is an individual who performs the task of eliciting knowledge. In some cases the Facilitator and the Analyst may be the same individual.\\
			\hline
		\end{tabular}
		\label{Tabledef}
	\end{table}

	\subsection{Challenges in the Elicitation Process} \label{Challenges}
	\underline{\textbf{Cognitive Biases}} \\
	
	\cite{tversky1974} pointed out that natural thought processes may create inaccuracies in the obtained priors by introducing cognitive biases not observable by the analyst. As these cognitive biases are present in the mind of the expert it is often complicated to adjust for them in the elicited distribution. While many methods have been proposed to mitigate these biases, it could be argued that the biases can never be completely eliminated. An overview of key cognitive biases and bias reduction methods for prior elicitation are outlined in Section \ref{CB}. \\

	\underline{\textbf{Using Multiple Experts to Elicit Priors}} \label{IntroME} \\
	
	It is good practice to use multiple experts to elicit a prior distribution \citep{o2019expert,williams2021comparison}. Doing so may ensure that the elicited distribution is better aligned with the entire field of interest, and that individual biases present in one expert are reduced. The process of obtaining a prior distribution that aggregates the inputs from multiple experts provides a new set of challenges. Two main approaches to aggregating the prior beliefs of multiple experts have been proposed. Each approach has its benefits and drawbacks; these are discussed in Section \ref{ME}\\

	\underline{\textbf{Dealing with Little or No Updating Data}} \label{LND}
	
	Many applications exhibit non-trivial complications. These include applications where there is unlikely to be much data available to update prior beliefs as needed when using Bayes' theorem. The absence of data for updating can occur because the data collection mechanism is too complex or expensive or because the data pertains to an occurrence of a highly undesirable event (e.g., a major accident or a security threat, etc.). In such situations, the decision making will be predominantly based on the elicited prior distributions since the data will have very little or no influence. A simple example is a problem in the security field:
	
	\begin{lemma}\label{example2}
		A company wishes to predict whether or not their new security system will fail, $Y \in \{1,0\}$.  Therefore $Y \sim Bernoulli(\theta)$ where $\theta$ is the probability of a system fail. For inference, data on how well the system performs is not present because that would mean security threats have already happened. If the security threat was to materialize, this will result in a single observation on $Y$. While $p(\theta|y)$ can now be computed, it is likely to be only slightly different to $p(\theta)$. Further, in practice, such a breach of the security system will likely result in changes being made so as to prevent such occurrence in the future. This may mean that the circumstances on which the original prior was elicited are no longer present and therefore eliciting a fresh set of prior distributions could be considered more appropriate rather than using the posterior obtained on the original prior.
	\end{lemma}
	
\begin{figure}[bt]
		\centering
		\includegraphics[width=6cm]{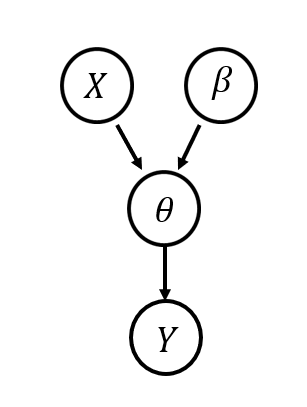}
		\caption{Graph showing the model structure.}\label{fig:ms}
	\end{figure}

	Another important facet of such applications is that while it may not be possible to update the prior beliefs on $\theta,$ there likely exists some background/ additional information, $x$, that may help in eliciting $p(\theta)$. Hence, $p(\theta)$ can be written as a function of $x$ and other model parameters $\beta$. That is, $p(\theta) = g(x,\beta)$ (see Figure \ref{fig:ms}). This could be thought of as a standard regression problem; however, there may be situations where background information, $x$, is disjointed and can contain different data types; such as image data, text data and expert opinion. Also, some of this information may only be vaguely relevant. Therefore, treating this example as an elicitation process may seem more reasonable. It is important to note that this type of application does not seem to be discussed much in the prior elicitation literature. Yet, it represents a practical decision theoretic problem whose solution may rely on the elicited probabilities.\\ 
	
	\noindent \textbf{Example 2 (contd.).} \textit{The company brings in experts to provide information on system failure. The company provides the experts with all information on their system (e.g., system type, location,...). The experts are able to give opinion on system failure based on the information provided and historical information they have obtained from other establishments and different security systems, $x$. Because each establishment is unique, what happened elsewhere may only be vaguely relevant to this company. The experts would aim to utilise all this related information to elicit a prior distribution on system failure.}\\

	\subsection{Motivation}
	This paper outlines methods used for eliciting informative priors and the benefits and drawbacks of these methods. Review articles in this field tend to focus on direct interrogation methods (Section \ref{DI}) (see for example \cite{hanea2021uncertainty,galway2007subjective,o2006uncertain} and \cite{jenkinson2005elicitation}), without exploring the other available methods presented in the literature. Other review articles  focus on "good practice" and answering questions which may arise in the overall elicitation process but do not expand on actual methods for eliciting priors \citep{stefan2020practical}.To the best of our knowledge, there has been no work to date that outlines all of the available approaches and provides an overview of the current state of art. In this paper we aim to provide such an overview and also to identify some of the gaps in the current literature.\\
	
	\subsection{Outline}	
	
	The elicitation methods are classified into three approaches: Interrogation Methods (Section \ref{IBM}), Graphical/Visual Methods (Section \ref{GVM}), and Historical Information Methods (Section \ref{OM}).  Interrogation methods are elicitation methods that involve asking an expert questions to obtain a probability distribution. Graphical/Visual Methods  involve plots of distributions or data that the expert can visualise and thus compare with their individual knowledge. Methods that are labelled Historic Information Methods are those that may not rely on expert input and instead utilise information from past research studies. An overview of cognitive biases is presented in Section \ref{CB}. Techniques for combining information from multiple experts are reviewed in Section \ref{ME}.  The paper concludes in Section \ref{sac}, by discussing some of the persistent challenges of informative prior elicitation and proposing further research paths.\\
	
	\section{Cognitive Biases in Expert Knowledge Elicitation} \label{CB}
	The following are considered key cognitive biases that may be present in the expert elicitation process (for a comprehensive list of issues which arise in human judgement see \cite{hogarth1987}).\\
	
	\begin{itemize}
		\item\textit{Judgement by Representativeness}; this may be present with questions such as ``What is the probability that an object A belongs to a class B?" \citep{garthwaite2005}. Often with these questions, the expert will focus on the conditional probability and completely ignore the unconditional probabilities. \cite{tversky1974} outline a clear example of Judgement of Representativeness in assigning the probability of an individual having a certain job, when the expert knows the individual's personality. The expert ignores the number of people in the population in the jobs and instead relies solely on the conditional probability, that is, they judge the probability of an individual having a certain job given the fit between the individual's personality and the job  \citep{tversky1974}.\\
		
		\item \textit{Judgement by Availability}; this is when an event is given a higher probability based solely on the fact that it occurred more recently for the expert \citep{tversky1974}, so they are able to easily bring it to mind. Considering Example \ref{example1}, an expert may have recently come across a case where a patient developed blood clots from the new drug. Even though the example may be rare, because it is fresh in their mind, the expert may believe that the risk of developing blood clots is higher, in turn, influencing the elicited prior.\\
		
		\item \textit{Anchoring and Adjustment}; here an expert is given some value (the anchor) and adjusts it to achieve what they think is the correct value. Experiments have shown that an expert who starts with a higher anchor is more likely to give higher estimates than experts who start with lower anchors \citep{tversky1974}. In Example \ref{example1} if the facilitator states a value for the mean of the probability of a patient obtaining blood clots, the expert may then adjust this value to what they think is suitable, producing results that are inaccurate due to the initial anchor value presented to them.\\
		
		\item \textit{Over Confidence} of the expert; Overconfidence is easily observed in studies where experts were asked for an interval \citep{soll2004overconfidence}. For example, where experts were asked for 95\% probability intervals, it was found that as little as 65\% of those intervals actually contained the true values \citep{o2019expert}. Thus, the expert often narrows down their judgment of uncertainty, displaying overconfidence.
		\cite{klayman1999overconfidence} gives an overview and comparison of methods which help evaluate an experts overconfidence, which may be helpful to researchers.\\
		
		\item \textit{Range Frequency}; this bias is seen when an expert is asked to assign the probability of each category and they evenly assign probabilities between categories \citep{parducci1963range,o2019expert}. This means that categories believed to be more likely are given less probability than required and others are given more, creating a significant bias . For Example \ref{example1}, an example could be when asking for the probabilities for a male developing blood clots and a female developing blood clots; the expert assigns equal probability to each group.\\

		\item \textit{Expert Fatigue}. When an expert is required to take in and store copious amounts of information or complex information to complete the elicitation process, the process can become cognitively tiresome and time consuming for the expert \citep{werner2017expert}. Analysts can reduce this burden by simplifying the models parameters, by splitting the task up into different parts and by explaining complex quantities in layman's terms for the expert \citep{werner2017expert}. All the while, analysts must keep in mind the trade-off between easing expert fatigue and still having models that represent the real world complexities. Expert fatigue varies from expert to expert and is heavily dependent on the level of expertise needed for the task \citep{barons2021balancing}. \\
	\end{itemize}

	Because cognitive biases are present in any human thought process, one goal of the expert elicitation methods should be to reduce these biases as much as possible.   These cognitive biases can be reduced by changing the types of questions asked \citep{meyer2001eliciting}. However, the way questions are asked may still affect the elicited prior.  Another approach to reducing cognitive biases  is a calibration technique, first introduced by \cite{lindley1982improvement}. The goal of this technique is to calibrate the expert's opinion based on the biases that they have shown in past opinions they have provided \citep{lichtenstein1977calibration,perala2020calibrating}. The technique involves using information from past estimates given by the expert where the true value is now known. \cite{perala2020calibrating} use hierarchical Gaussian processes to model the expert's biases from their past estimates. The models are trained on the calibration data (historical estimates and the true values) and then used to correct for bias in the new estimates given by the expert \citep{perala2020calibrating}. Although a promising technique, the obvious disadvantage is, when calibration data is limited or non-existent, that this technique becomes impossible.\\ 
	Others have suggested that the best way to address cognitive biases is to give the expert feedback and make them accountable for there responses \citep{russo1992managing}. Training them to give more appropriate responses in the future \citep{lichtenstein1977calibration}.
	Cognitive biases have been outlined in detail in terms of expert prior elicitation in \citep{garthwaite2005,o2019expert,stefan2020practical}. \\
	Methods to elicit an informative prior, discussed in Sections \ref{IBM},\ref{GVM} and \ref{OM}, aim to reduce such biases; however, the biases above may still occur in each method.  \\

	\section{Interrogation Methods} \label{IBM}
	Interrogation methods are those in which an expert is interviewed to obtain knowledge on a parameter of interest \citep{chesley1975elicitation}. \\
	\subsection{Direct Interrogation} \label{DI}
	Distribution or probability based questioning is the most common form of prior elicitation \citep{galway2007subjective,jenkinson2005elicitation}. Classified as direct interrogation methods, these techniques use a questionnaire or interview with an expert, with questions that are specifically related to probabilities and/or distributions.  \cite{johnson2010methods} provide a systematic review of papers that involve direct interrogation methods for prior elicitation, outlining specific papers and the questions used. \cite{meyer2001eliciting} outline procedures for researchers to create their own questions for their specific task and refine them where necessary, taking into account potential biases that may arise.  Techniques for direct interrogation include:\\
	
	\begin{itemize}
		\item \textbf{Questions on Probabilities}: Asking questions on probabilities that could directly link to the Cumulative Distribution Function (CDF) or Probability Density Function (PDF) \citep{o2006uncertain}. For instance, when eliciting a prior for Example \ref{example1}, a possible question could be ``What is the probability that the proportion of those who get blood clots is less than or equal to 0.3?", i.e., $P(Y\leq0.3)$. Once a few points have been collected, the researcher can obtain an outline of the CDF or PDF.\\
		
		\item \textbf{Questions on Distribution Quantiles}: Asking questions on the quantiles of the CDF or PDF \citep{winkler1967assessment}. In the case of Example \ref{example1}, a possible question could be ``At what value is the probability of a patient obtaining blood clots  equally likely to be less than or greater than that value?" ( i.e., estimate the median).\\

	\end{itemize}
	
	Under each of theses techniques, the facilitator can either ask the expert to give a value for the variable given a fixed probability or ask the expert to give a probability for a fixed value or ask questions where to expert is expected to give both probability and value \citep{spetzler1972probability}.
	Some analysts prefer to ask questions directly on the parameter of interest. This is known as \textit{Structural Elicitation} \citep{kadane1998experiences}, this should not be confused with \textit{Structured Elicitation} which is a step-by-step process of elicitation which may contain \textit{Structural Elicitation}. \cite{o2006uncertain} (Chapter 5) give an extensive review of such techniques, detailing acceptable methods and how to use the expert's responses to form a probability distribution (also summarised in  \cite{garthwaite2005}). Tools such as SHELF \citep{oakley2010shelf} and MATCH  \citep{morris2014web} are suitable for forming distributions from this line of questioning. These tools also provide instant visual feedback of the expert's opinions in the forms of fitting a distribution which has been found to be beneficial to the elicitation process \citep{o2006uncertain,johnson2010methods}. Techniques which directly involve graphical and visual components are discussed separately in Section \ref{GVM} of this paper. \\
	
	\cite{kadane1980interactive} argued that in some cases parameters are arduous to think about contextually, which can create difficulties in the elicitation process; as the expert will find it difficult to form a belief on the parameter's behaviour. Therefore, questions should be asked on observable values, i.e., the predictive distribution.  This approach is known as \textit{Predictive Elicitation}.  While some experts may find it hard to understand the real-world applications of model parameters, experts should, by definition, have copious knowledge on the field of expertise data.  Predictive Elicitation has been explored further in \cite{kadane1998experiences,akbarov2009probability} and \cite{hartmann2020flexible}, with \cite{kadane1998experiences} providing a comparison of Structural and Predictive Elicitation, showing examples of when each technique is appropriate.  \cite{kadane1998experiences} emphasise that creating a predictive elicitation process is task specific and must be adapted for each task undertaken by the analyst. Predictive elicitation techniques are also more complex and time consuming to perform than structural techniques \citep{kadane1998experiences}. It is worth noting that these techniques essentially treat prior elicitation as an inference problem. Consider a basic example for a regression model, a facilitator would ask the expert about their uncertainty around the dependent variable given different values of the response variable. Once the uncertainty is captured, the analyst would then have to infer the range of values for the model parameters and their prior distributions that would be consistent with the expert's elicitation \citep{kadane1998experiences}.\\ 
	
	Applying direct interrogation methods to Examples \ref{example1} and \ref{example2} may be appropriate if the expert has good comprehension of statistical concepts needed.  But even if they do, common cognitive biases, such as \textit{Expert Fatigue}, \textit{Judgement by Representativeness}, \textit{Judgement by Availability}, \textit{Anchoring and Adjustment}, \textit{Over Confidence} and \textit{Range Frequency}, may still affect the process. This is when calibration techniques, as discussed in Section \ref{CB} may be of use.\\
	Direct Interrogation Methods also have the additional obstacle of first requiring the facilitator to teach the expert to understand statistical concepts, which can be complex. Even once the elicitation process has started, the expert may still not understand key concepts needed, nor be able to apply them to form an accurate probability statement. \cite{kadane1998experiences} state \textit{"The goal of elicitation, as we see it, is to make it as easy as possible for subject-matter experts to tell us what they believe, in probabilistic terms, while reducing how much they need to know about probability theory to do so"}. Hence, other types of interrogation methods exist that can reduce the statistical knowledge required to complete the task of eliciting a prior from experts, discussed next in Section \ref{ITM}. Some other relevant methods are discussed in Sections \ref{GVM} and \ref{OM}. \\

	\subsection{Indirect Interrogation} \label{ITM}
	Probability-based questioning is not the only form of questioning in interrogation techniques. There are  approaches to form distributions from other types of questioning, known as indirect interrogation techniques. \\
	
	\begin{itemize}

		\item \textbf{Betting method:} In this method, a series of bets are placed to form a distribution \citep{winkler1967quantification}. The process starts by presenting the expert with two bets and asking the expert to select the bet based on the event they believe is more likely to occur. An example from \cite{winkler1967quantification} is as follows:
		Let \textbf{Bet One} be ``\textit{Win \$A if Event E occurs, lose \$B if E doesn't occur}" and \textbf{Bet Two} be ``\textit{Win \$B if E doesn't occur, lose \$A if E occurs}". The expected values of each will be \textbf{Bet One}: $Ap - B(1-p)$, \textbf{Bet Two}: $B(1-p) - Ap$. If the expert chooses Bet One , then for this expert $Ap - B(1-p) \geq B(1-p) - Ap \Rightarrow p \geq B/(A+B) $. By letting Event, E, be any combination on the real line, analysts can elicit information on a probability distribution. Another approach is to use a probability wheel, split into two colours, where an expert can select between two scenarios; either they win $\$X$ if the probability wheel lands in a grey area or they win $\$X$ if an event occurs \citep{spetzler1972probability}. The facilitator then changes the area of the grey area and/or the event specifications to present a new question to the expert. This is repeated until the expert finds both situations equally likely \citep{spetzler1972probability,abbas2008comparison}.
		The number of bets to form an appropriate distribution could be relatively large, causing expert fatigue; to avoid this, betting can also be used alongside direct interrogation methods to check the certainty of assessed probabilities \citep{winkler1967quantification}.  \\
		Using the betting method for both Example \ref{example1} and \ref{example2} may require extensive thought and time. Setting up suitable bets will be the first obstacle, but then the analysts must create enough bets to form a more complex distribution, which can be time consuming. Betting situations may prove more useful in the case where an analyst needs to elicit an individual probability rather than a complete distribution.\\
		
		\item \textbf{Pair-wise Comparison:} In this approach, experts compare pairs of categories and quantify their beliefs regarding which category is more likely. This action is carried out for each possible pair in the study. An Analytic Hierarchy Process (AHP) \citep{saaty1987analytic,saaty1980ahp} is a pair-wise comparison process which can also be used for prior elicitation \citep{cagno2000using}.  For comparisons, experts are asked to give a number on a scale (from 1 - 9) based on which event they believe is more likely to occur. Using Example \ref{example2}, an AHP approach could be used to obtain prior distributions for different locations of sites where the security system is used. For example, let building location be the set \{``Country", ``City Centre", ``Suburbia", ``Industrial"\}. The expert can then compare which system is more likely to fail for each pair of building locations. This forms an opinion matrix, as shown in Table \ref{TableAHP}.\\
		\begin{table}[h]
			\caption{Opinion Matrix for the AHP Method for Building Location in Example 2}
			\centering
			\begin{tabular}{l| c c c c}
				\hline
				Building Location & Country & City Centre & Suburbia & Industrial \\\hline
				Country  & 1  & 1/3 & 1/3 &  1/7\\\hline
				City Centre & 3 & 1 & 1 & 1/5\\\hline
				Suburbia & 3 & 1 & 1 & 1/5 \\\hline
				Industrial & 7 & 5 & 5 & 1\\
				\hline
			\end{tabular}
			\label{TableAHP}
		\end{table}

		Line four in the opinion matrix reads \citep{cagno2000using} that the expert believes buildings in Industrial areas  are very strongly more probable to have security failure than buildings in Country areas and strongly more probable to have security failure than buildings in the City Centre and Suburbia areas. By taking the maximum eigenvalue of the opinion matrix above  and calculating the associated eigenvector, the analyst can obtain the vector of weights of the building areas \citep{saaty1987analytic,saaty1980ahp}, where the weights are the expert's security failure likelihood (or propensity-to-failure in \cite{cagno2000using}) for each building location. \cite{cagno2000using} then calculated the mean and standard deviation of the expert's opinions and used these to obtain the parameters of a Gamma distribution for each class.\\
		Although there may be some use for this method in the case of Example \ref{example2} (where there is potential for sub-classes and experts have observed examples from each class), for Example \ref{example1}, reducing the case down to specific sub-classes may not necessarily be possible. This method obviously becomes impractical when there are no sub-classes in the research task.
		\\
		
		\item \textbf{Ranking/Rating Method:} in this method, experts are asked to rank or give a rating on the likelihood of events presented to them \citep{eckenrode1965weighting,edwards1994smarts}.  \cite{wang2013expert} give an appropriate example of this method in terms of decision analysis.  In the application of understanding terrorist attacks, they asked intelligence officers to rank the attractiveness of selected potential targets. They then used probabilistic inversion and Bayesian density estimation to elicit the distributions \citep{wang2013expert}. This method can also be used for prior elicitation under time pressure \citep{jaspersen2015probability}.  A ranking method may be hard to implement for Example \ref{example1}, where the analyst may find it difficult to split the problem into sub-groups that would be rank-able. For Example \ref{example2}, this method may work quite well. The analyst has the additional information on the property so they could come up with other properties that have different attributes to the property of interest and ask the expert to rank a list of properties.\\
		
	\end{itemize}
	
	These techniques are designed to avoid having to teach the experts about statistical concepts, which can be a drawback of direct interrogation methods \citep{wang2013expert}. They can also simplify the process, however, analysts need a strong grasp on the subject matter to be able to form a suitable set up for elicitation (e.g., forming sub-groups and different attributes which may effect the elicited distribution). Indirect interrogation methods also clearly remove the cognitive biases of \textit{Judgement of Representativeness} and \textit{Range Frequency}, by not asking questions relating to probabilities. However, other cognitive biases may still be present. 
	\\
	
	\section{Graphical/Visual Methods:}\label{GVM}
	Graphical/Visual Methods are methods which involve graphical representations of the probability distribution and/or data which may be used to form a distribution.\\
	
	\begin{itemize}
		\item \textbf{Trial Roulette method}: First established by  \cite{gore1987biostatistics}, this method involves a graphical representation of the parameter space of the parameter of interest being split into subsections. The expert then assigns blocks to the subsections, essentially building their own probability distribution. The trial roulette method has been used in practice in \cite{diamond2014expert}. \\
		In Example \ref{example1},  the online tool MATCH \citep{morris2014web} can be used to elicit the expert's belief via the Trial Roulette Method. The MATCH tool for the Trial Roulette method is shown in Figure \ref{fig:mr}. In the particular example scenario considered in Figure \ref{fig:mr}, the expert has  placed the majority of the blocks towards the zero end of the scale, meaning they believe the probability of someone developing blood clots is small. They still, however, have some uncertainty, having placed blocks as far out as 0.5 to cover this uncertainty regarding the location of the true value. In this scenario, the tool has found the $Gamma(1.03,6.25)$ distribution to be an appropriate fit for the expert's distribution. The user may change this distribution if desired, and select (say) a Scaled-Beta which may be considered to be more appropriate given its support and conjugacy to the Bernoulli likelihood.\\
		Although the Trial Roulette method is a great visual tool, it still requires the expert to have sufficient statistical knowledge to be able to place the blocks to form an appropriate distribution. Also, the user needs to understand probability distributions to avoid inaccuracies in the default selections. For instance, in the example discussed above,  conceptually, a Gamma distribution is not appropriate for proportions as it can take values greater than one.  
		\\ 
		
		\begin{figure}[h]
			\centering
			\includegraphics[width=1\linewidth]{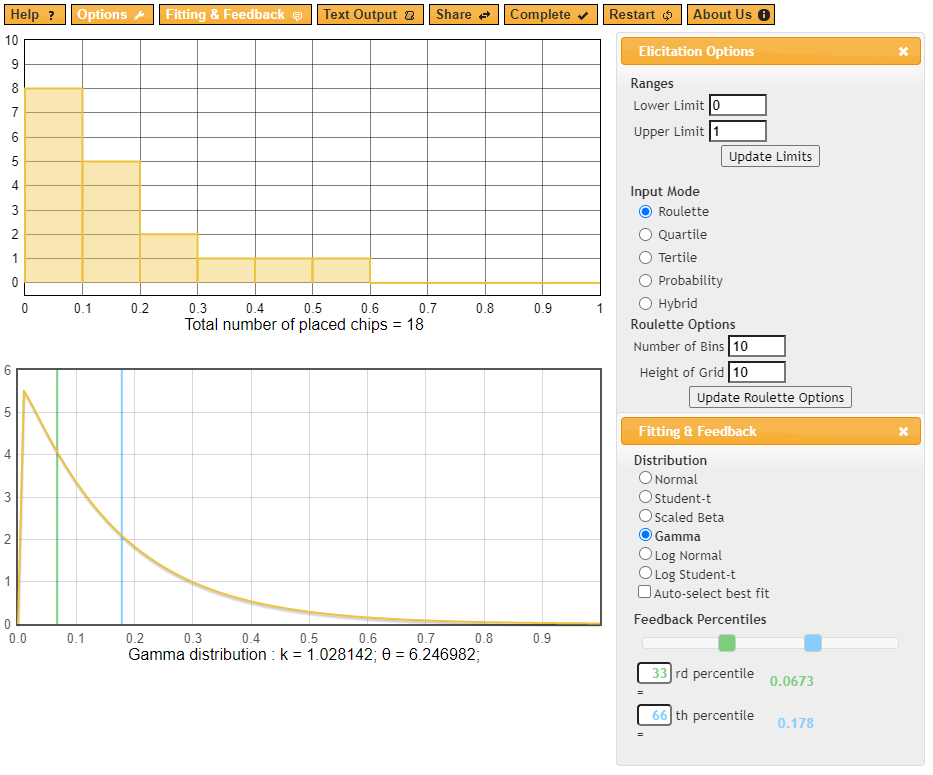}
			\caption{Online Tool MATCH displaying the Trial Roulette Method}\label{fig:mr}
		\end{figure}
		
		\item \textbf{Graphical Prior Elicitation}: \cite{casement2018graphical} outline a simple graphical tool for prior elicitation in univariate models. This tool initially teaches the expert about the different distribution types and what effect the distribution parameters have on the them. Explaining this method with Example \ref{example1}, the user begins by selecting an initial model distribution; the distribution on $Y$, a Bernoulli distribution. The tool automatically selects a conjugate prior distribution on the parameter, a Beta Distribution. The user will then select the expected number of successes out of 100, automatically generating plots of what  the number of successes vs. failures would look like, see Figure \ref{fig:get}. Changing the number of successes, the expert can then see how the resulting plots will change. This can be thought of as a Predictive Elicitation technique as the user is observing plots of the data, not the parameter. After training the expert, the tool then displays a collection of sample data histograms of which the expert must select those that are similar to what they believe the data would be. From there, the prior distribution is formed from the selected histograms. \cite{casement2018graphical} suggest this method simplifies the statistical knowledge required to produce a prior distribution, significantly reduces the time of training the expert and the processing time of forming a prior distribution. However, experts still need partial statistical understanding to complete the task of outputting a prior that is consistent with their beliefs. \\

		\begin{figure}
			\centering
			\includegraphics[width=1\linewidth]{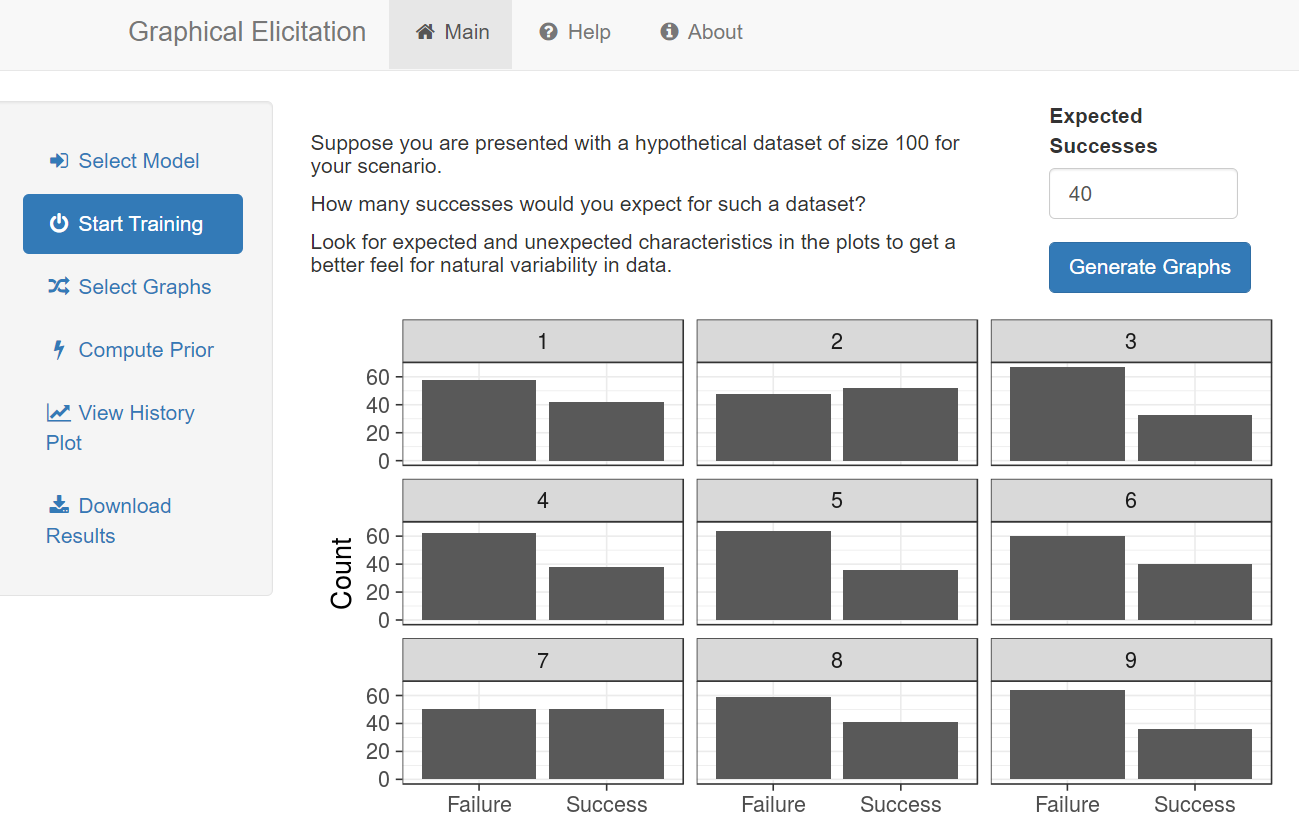}
			\caption{Training Step of the Graphical Elicitation Method.}\label{fig:get}
		\end{figure}
		
		\item \textbf{Interactive Excel Spreadsheet}: Starting with a plot of a selected prior distribution (for example, the user may have selected a Gaussian distribution), this method involves the expert moving the ``sliders", in the corresponding excel spreadsheet, to see the effect of changing the value of the parameters on the prior distribution \citep{jones2014prior}. This approach allows the expert to adjust the parameters until they find their ideal prior distribution. For the case where parameters are not ``intuitively meaningful" \citep{jones2014prior}, the interactive spreadsheet, used in this method can show  multiple plots of the observable quantity's distributions, allowing the user to see the effects of changes to these distributions when parameters of the original prior distribution are changed. This variation can be thought of as a Predictive Elicitation form of this method. \\
		This spreadsheet based method is similar to the Graphical Prior Elicitation Method in that it allows for experts to first learn the effects of distribution parameters on the given distribution. Also, like the Graphical Prior Elicitation Method, experts still need to have comprehensive knowledge of statistical concepts to be able to understand the distributions themselves, including knowledge of which distribution to initially select. \cite{jones2014prior} provide example spreadsheets in their supplementary materials and give an overview of different applications where this technique has been applied.\\
		
		\item \textbf{Computer Simulated Data}: \cite{thomas2020probabilistic} introduce a method which involves a plot of simulated data. They illustrate two techniques; in the first  technique, the expert observes a plot of simulated data and responds as to whether or not this data could be from a real data set. The second is where pairs of simulated data are compared to one another. Again, this can be thought of as a Predictive Elicitation technique. From there, the expert's responses are used to train a Gaussian Process (GP) classifier that  captures the expert's belief \citep{thomas2020probabilistic}. Exploring the first technique further, an expert is shown a simulation $Y_\theta \sim p(Y|\theta, M)$, conditional on model parameters $\theta$, drawn from model $M$. The response from the expert is binary, $z_E = 1$ or $0$. These responses are then used to train a GP classifier, $C$,  to model $p(z|\theta)$ \citep{thomas2020probabilistic}. Using Bayes theorem, Thomas et al. describe how they can estimate $p(\theta| z=1, C, z_E, M)$.
		GP classifiers work by placing a GP prior over a latent function, $f$, and then applying a logistic function, $\sigma$, to $f$ to obtain a prior for class probabilities, $\pi$, that is $\pi(x) = \sigma(f(x))$. \\
		This method is similar to Graphical Elicitation. The main difference is the selection of the prior distribution type, as this method does not assume a conjugate prior. It is note worthy that in this method, the expert does not need a strong knowledge of statistical concepts to obtain an accurate prior distribution; as they should have a strong understanding of subject matter data.\\
	\end{itemize}
	
	Although graphical methods are helpful in providing instant visual feedback to the expert, and, in some cases, simplifying the statistical knowledge required to perform the task, they are still subject to the cognitive biases discussed in Section \ref{CB}. The Computer Simulated Data method and the Graphical Elicitation Method, are interesting approaches, because they aim to generate a distribution by modelling the decisions of the expert. This may be an interesting topic for further research.  \\

	\section{Using Historical Information} \label{OM} 
	Methods discussed thus far aim to elicit the uncertainty that the expert has into a probability distribution. Thus, they could be considered to be similar to a standard translation tool, with the objective of translating the expert's knowledge, including their quantified uncertainty, into a probability distribution. 
	In practice, a standard translation tool that translates a text from one language into another, requires the text to be available in the first place. In the same way, the prior elicitation methods discussed so far require the expert to quantify their uncertainty first.
	However, there are applications where relevant historical data exists, and the analyst may prefer to quantify uncertainty using such data instead. In this section, methods that elicit prior distributions using available past research are reviewed. The main theme of these methods is that they require minimal human intervention to obtain a prior distribution.\\
	
	\begin{itemize}
		\item \textbf{Utilising Historical Research Posteriors:} One way to obtain a prior without expert intervention is from similar historical research that provides a posterior distribution. This posterior distribution can then be used as a prior distribution in a new study \citep{press2009subjective}. However, many similar studies will not provide a concrete posterior distribution, but may have similar data.\\
		
		\item \textbf{Power Prior:} The power prior \citep{ibrahim2000power} provides a way to obtain a prior from historical data, $D_0 = (n_0, y_0, X_0)$, where $n_0$ denotes the sample size, $y_0$ denotes the response vector and $X_0$ denotes the matrix of the covariates.  In a basic form, the power prior is the likelihood of the historical data, $L(D_0|\theta)$, to the power of a scalar, $a_0$, multiplied by an initial prior $\pi_0(\theta|c_0)$, 
		\begin{equation}\label{eq:powerprior}
		\pi( \theta | D_0, a_0) \propto L(D_0|\theta)^{a_0} \pi_0(\theta|c_0).
		\end{equation}
		
		The scalar, $a_0$, quantifies the uncertainty in the data, $D_0$. It allows the analyst to control how much the historical data influences the prior. The initial prior parameter, $c_0$, allows the analyst to control the influence of the initial prior on the prior, i.e, how much information from the initial prior is influencing the power prior. As selection of these parameters can change the final power prior, analysts should take careful consideration when deciding these values \citep{ibrahim2000power}. Other variations of the power prior \citep{ibrahim2015power} may be of use to some practitioners, along with methods to form priors from not only historical data but also historical research that only provides summary statistics of the data \citep{chen1998heritability}.\\
		
	\end{itemize}
	
	These methods help mitigate cognitive biases as they do not rely on the human thought processes to quantify uncertainty.  For instance, in Example \ref{example1}, before releasing the drug for human tests, the drug must be taken through animal trials \citep{turner2010new}. The trials done on animals could be relevant historical data, which could help formulate a prior distribution for human patients. Similarly, data from \textit{Phase I} clinical trials can be used to formulate a prior distribution for \textit{Phase II/III} clinical trials, and so on. It could be argued, however, that biases  may have arisen in the creation of the historical model and data selection.\\
	The disadvantage of these techniques is that they require access to similar research and/or historical data. Many subject fields may not have similar studies to the current research task and/or access to historical data is restricted, rendering these methods futile. The problem of not having access to historical data is relevant to Example \ref{example2}. Although it may be possible to find information on a similar security failure in a similar organisation, the amount of data needed to form an appropriate distribution may be lacking. Also, as stated in Section \ref{LND}, once a security threat has been observed, the system may be altered, making the new system of the similar organisation different to the original system in any potential historical data.\\

	\section{Eliciting Priors from Multiple Experts}\label{ME} 
	The methods outlined in Sections \ref{IBM}-\ref{GVM} give an overview of appropriate techniques to elicit prior distributions from an individual expert. However, as outlined in Section \ref{IntroME}, analysts should use multiple experts to form a probability distribution representative of the whole field of research \citep{o2019expert,williams2021comparison}.
	When selecting an appropriate method, the analyst must trade off simplicity of method and likely accuracy of the resulting prior.Extensive reviews of methods for combining multiple expert beliefs can be found in \cite{ rougier2013risk,winkler1968consensus, o2006uncertain} and \cite{clemen1999combining}. A comparison of some aggregation techniques can be found in \cite{williams2021comparison}. \\
	An overview of methods to combine information from multiple experts follows.\\

	\textbf{Behavioural Aggregation Methods:} 
	These methods involve groups of experts discussing their beliefs and, in some cases, reaching a consensus. \\
	\begin{itemize}
		\item \textbf{Delphi Method} \citep{helmer1967analysis}: This method most commonly starts with a questionnaire that is sent out to all the selected experts. Opinions from the initial questionnaire are then sent out to the same experts along with a new questionnaire, thus giving experts the option to update their own opinions based on their peers responses \citep{niederberger2020delphi}. This process may be repeated until consensus is reached \citep{rougier2013risk}. \\
		
		\item \textbf{Interactive Group Methods}: In these methods, experts and a facilitator meet to discuss and come to a consensus on a probability distribution \citep{rougier2013risk,o2006uncertain}. The SHELF method \citep{oakley2010shelf} is a very common interactive group method. It is a framework for facilitators to follow that allows experts to reach a consensus. \\
	\end{itemize}
	
	Although behavioural aggregation may appear to be an ideal way to aggregate expert opinion because it allows for shared expertise, the goal of reaching a `consensus' can be problematic. This is because social psychological factors can distort the process of reaching consensus \citep{baron2005so}. Interactive groups are particularly vulnerable to group think, seniority, titles or social hierarchy. For example, people who are high on personality characteristics such as social dominance or who are considered by other experts in the group to have more expertise (e.g., more citations) can influence a consensus prior so that it resembles the dominant individual's prior and doesn't actually capture the variability in certainty across the group. This problem can be overcome in some variants of the Delphi method where experts are anonymous to each other (e.g., online methods). \\
	Over-confidence may also be an issue for consensus priors from behavioural aggregation. In an evaluation of methods of aggregation, \citep{williams2021comparison} compared the SHELF method with mathematical methods. They speculated that over-confidence may have been present in the SHELF method because it yielded narrower distributions. Moreover, not only is forming a consensus difficult, but the whole process of behavioural aggregation can take time \citep{williams2021comparison}. For the interactive group technique there can be difficulty in aligning work schedules, deciding on an optimal time to meet and the discussion itself. For the Delphi Method, there may be significant wait time during the process of having to send responses back and forth to expert group members. All of these issues may be resolved through mathematical aggregation discussed below.\\

	\textbf{Mathematical Aggregation Methods:} These methods involve mathematical techniques to combine individual expert priors elicited by methods in the sections above.\\
	
	\begin{itemize}
		\item \textbf{Bayesian Method}: This approach treats each expert's prior as new data. The analyst updates their prior with the ``new data". The resulting posterior consists of the combined prior beliefs \citep{morris1974decision,lindley1983reconciliation,albert2012combining}. This process can be complex and time consuming \citep{o2006uncertain}.\\
		
		\item \textbf{Opinion Pooling}: These methods form a consensus distribution $f(\theta)$ as a function of the individual distributions ($f_1(\theta)$, $f_2(\theta)$, ...) \citep{o2006uncertain,genest1986combining}. A simple example of this is Linear Pooling where the consensus distribution is equal to the weighted sum of the individual distributions. That is,
		\begin{equation}
		f(\theta) = \sum_{i=1}^{n} w_i f_i(\theta)
		\end{equation}
		where $n$ is the number of experts. Other common pooling methods can be found in \cite{o2006uncertain}. Key methods for finding weights are listed below:
		\begin{itemize}
			\item Equal Weighting: Give each distribution the same weight, $1/n$.\\
			\item Self-Weighting: Each expert gives themselves the appropriate weighting. \\
			\item Expertise/Performance weighting: Not all experts may be considered equal. Some methods suggest weighting experts based on their expertise or performance in their field \citep{winkler1968consensus, o2006uncertain,lindley1986reliability}. As an example, an analyst may decide that the longer someone has been working in the field, the higher the weight their prior receives. Cooke's Method \citep{cooke1991experts}, is a performance weighting method that gives the experts an added task (a smaller elicitation process where the subject matter parameter is known) to assess performance. An application of Cooke's method is outlined in \cite{aspinall2010route}.\\

			\item AHP Weighting: Also encountered in Section \ref{IBM}, an Analytic Hierarchy Process \citep{saaty1987analytic} is a process where an analyst can compare all solutions against a decision criterion \citep{syed2020novel}. In the case of weighting expert priors, \cite{syed2020novel} used three criteria for expert assessment: ``years of experience", ``number of observed failures", and ``level of training". The weights obtained from the process are used in two ways: firstly, to create a shortlist of potential experts and secondly, as the weights for the aggregation process. \\
		\end{itemize}

	\end{itemize}
	Selection of weights is a very important aspect of mathematical aggregation. Selection methods like equal weighting and self weighting are the easiest methods to apply. However, equal weighting treats all experts as being equally knowledgeable on the subject matter, which is generally not applicable to real life. Self-weighting methods can also be inaccurate. It has been found in the past that with self-weighting, women ranked their expertise lower than men \citep{brockhoff1975iv}. This is still the case today where men tend to be more self-promoting than women \citep{exley2019gender}. Also, with overconfidence already being a bias present in expert elicitation, it may become particularly detrimental in this selection method. In contrast, methods based on performance, or some other objective criterion seem appropriate, as they take into account the level of expertise of any given expert in a less subjective manner. For an overview of selecting weights, in mathematical aggregation, see \cite{rougier2013risk}.\\

	\textbf{Hybrid Methods:}\\
	To utilise the benefits of behavioural aggregation and mitigate the drawbacks, there have been hybrid methods created which incorporate both behavioural and mathematical aggregation techniques \citep{ferrell1994discrete}. One such protocol is the IDEA protocol \citep{hanea2018classical}. "It encourages experts to \textbf{I}nvestigate and estimate individual first round responses, \textbf{D}iscuss, \textbf{E}stimate second round responses, following which judgements are combined using mathematical \textbf{A}ggregation" \citep{hanea2018classical}.\\

	\section{Summary and Conclusion}\label{sac}
	When information from the likelihood is limited, obtaining an accurate and  informative prior distribution is especially critical. This paper outlines some of the important approaches for eliciting prior distributions along with their merits and limitations. Each elicitation task is unique and so are the experts involved in that elicitation. Therefore, while it may be possible to find a suitable and appropriate prior elicitation approach for a given task, none of the approaches is arguably superior overall \citep{kadane1998experiences}. Moreover, there could be applications for which none of the existing methods are appropriate.  
	\\
	
	\subsection{Persistent Challenges}
	
	Where expert involvement is required, there will always be persistent challenges in obtaining an informative prior. This paper does not address the issues surrounding selecting experts. However, experts need to be highly knowledgeable in the relevant content area, and for a number of the methods, they also need to both understand the relevant statistical concepts, and be able to think in a statistically valid manner.\\
	Another related area of challenge is in minimising the influence of the common cognitive biases outlined in Section \ref{CB}. They are part of natural human thought processes that have proved adaptive in other contexts, particularly when making decisions under pressure. It is not realistic to expect that they can be eliminated. But a key goal of any expert elicitation method should be to reduce them as much as possible.\\ 
	Recall Example \ref{example2}, where the task is to elicit a prior distribution using related information $x$ that could be heterogeneous both in type and relevance. Common methods discussed in Section \ref{IBM} and \ref{GVM} look at getting information solely from an expert, methods in Section \ref{OM} look at solely getting information from the data. However, methods that enable the analyst to elicit a prior distribution using all of the heterogeneous relevant information available, as well as including expert knowledge, do not yet exist.\\
	Translating the uncertainty of an expert into a probability distribution may always remain a challenging problem. As discussed, there are many ways to do this. However,  the most suitable  method is usually task specific and should be thought of as such. There is no one superior way to quantify someone's uncertainty. The process used for each elicitation task should be selected based on the problem at hand, the resources available, the experts available, and their ability to quantify their uncertainty, as well as, the availability of any past research/data or relevant information.\\ 
	It is possible that an elicited prior distribution may not accurately capture the expert's beliefs and can be considered to be only an approximation at best. Importantly, it may never be possible to ascertain how accurately  a prior distribution reflects an expert's beliefs. Inaccurate elicitation of prior distributions may lead to inaccurate posterior inference and therefore, to inaccurate data analysis. This problem is exacerbated when the prior dominates the likelihood because of insufficient data.\\ 
	A possible solution to inaccurately elicited prior distributions is to implement a prior robustness analysis. This, typically involves defining a class of prior distributions which encompass the uncertainty around  the original prior \citep{basu1994variations}.
	Prior robustness analysis studies the sensitivity of the posterior distribution to the choice of prior distribution. Several prior robustness approaches have been developed, see 	\cite{insua2000} for a general introduction to prior robustness analysis and an overview of various approaches therein.  A recent approach to implement prior robustness, which is also straightforward to implement, uses distortion functions \citep{arias2016new} to form a \textit{distorted band} class of priors. One of the challenges in implementing a prior robustness analysis is in quantifying the uncertainty in the elicitation of the original prior. Methods to quantify such uncertainty have also been proposed. For the distorted band class of priors, \cite{joshi2017prior} suggest a simple interrogation approach that can determine the length of the distorted band class of priors to accurately quantify the expert's uncertainty around the original prior distribution.	\\
	
	In this manuscript the focus is on how an informative prior can be elicited, that is,  capturing expert's beliefs to form a probability distribution. This should not be confused with obtaining forecasts or estimates from an expert, which although can contain aspects of what is discussed in this paper, is not forming a probability distribution from their beliefs. For approaches of this nature, we suggest the reader look into the ``Good Judgement Project" and the research related to it \citep{tetlock2016superforecasting,ungar2012good,mellers2014psychological}.\\

	\subsection{Future Research}
	
	Persistent challenges open the way for research into new methods. Suggestions for potential research avenues are as follows. \\
	There is a need to find the most appropriate method for a given task. Therefore, the field of prior elicitation could benefit from having more research into comparisons of different methods. Although some research on comparing different methods is available, it often focuses only on one type of elicitation technique, see for example \cite{kadane1998experiences,abbas2008comparison}, or \cite{williams2021comparison}. More research into comparisons of a range of approaches will not only help practitioners gauge which method is more appropriate for their task, but it will allow for continual discussion on the topic of elicitation.    \\
	Research could also consider situations similar to the specific case of Example \ref{example2}, where data for the likelihood is unlikely to occur and when it does occur, it leads to system change. This implies that the initial prior needs to be reliable and, may need to be updated as the system changes. Redoing the same initial elicitation process could be a  solution, but this process will be time consuming and not always a viable option, so, other methods should be explored. Additionally, how to incorporate all the heterogeneous information available for the elicitation, would be an interesting topic for exploration.\\
	New prior elicitation techniques that focus on addressing some of the key challenges should also be researched further. One such technique could be on producing priors from modelling the expert decision making process. There do exist methods that model a simple decision making task, such as the graphical elicitation method discussed in Section \ref{GVM}, and  also methods briefly considered in Section \ref{ITM} (e.g. the ranking method). However, these methods rely on hypothetical decision making; that is, making decisions on circumstances that are not real. Research could instead focus on eliciting priors from real life decision making.\\
	Using real situations could lead to a more realistic prior than a hypothetical decision making task, because decisions made in real life carry far greater consequences than those in hypothetical situations, leading experts to make more effort to be accurate in the information they provide. It is worth noting that this decision making task may not be performed under the scope of prior elicitation; there may be past decisions that have been made that can be used to elicit priors from the decision makers. Using past decisions could help in creating a calibration technique, like those discussed in Section \ref{CB}, as the data may include the real life outcome after the expert's decision. This could allow for the analyst to compare the decision against the actual outcome and adjust for any biases that may arise in the decision making process, resulting in a fair prior. Another benefit of such a method is that the expert requires no statistical knowledge to elicit a prior, which can be a drawback of other methods. \\
	The role of prior robustness in prior elicitation should also be explored further. Focus could be directed towards the expert's uncertainty surrounding their prior and using different approaches to obtain this uncertainty to form bounds on the elicited prior distribution. One way could be to use a performance based criterion, like those discussed when combining multiple expert priors (Section \ref{ME}), in which experts who are considered ``more knowledgeable" in the field will have narrower bounds on their elicited prior.\\

	\bibliographystyle{unsrt}
	\bibliography{RParXiv}  

	
	
	
	

\end{document}